\documentclass[a4paper]{article}

\usepackage{INTERSPEECH2020}
\usepackage{tikz}
\usepackage{math,color}
\usepackage{amssymb}
\usepackage{balance}
\usepackage{url}
\title{Online Monaural Speech Enhancement Using Delayed Subband LSTM}
\name{Xiaofei Li$^1$ and Radu Horaud$^2$}
\address{
  $^1$Westlake University, Hangzhou, China\\
  $^2$Inria Grenoble Rh\^one-Alpes, Montbonnot Saint-Martin, France}
\email{lixiaofei@westlake.edu.cn, Radu.Horaud@inria.fr}

\begin{document}

\maketitle
\begin{abstract}

This paper proposes a delayed subband LSTM network for online monaural (single-channel) speech enhancement. The proposed method is developed in the short time Fourier transform (STFT) domain. Online processing requires frame-by-frame signal reception and processing. A paramount feature of the proposed method is that the same LSTM is used across frequencies, which drastically reduces the number of network parameters, the amount of training data and the computational burden. Training is performed in a subband manner: the input consists of one frequency, together with a few context frequencies. The network learns a speech-to-noise discriminative function relying on the signal stationarity and on the local spectral pattern, based on which it predicts a clean-speech mask at each frequency. To exploit future information, i.e. look-ahead, we propose an output-delayed subband architecture, which allows the unidirectional forward network to process a few future frames in addition to the current frame. We leverage the proposed method to participate to the DNS real-time speech enhancement challenge. Experiments with the DNS dataset show that the proposed method achieves better performance-measuring scores than the DNS baseline method, which learns the full-band spectra using a gated recurrent unit network.

\end{abstract}

\section{Introduction}
This paper addresses the problem of online single-channel speech enhancement/denoising, where 'online' means that the signal is received and processed frame by frame. Deep-learning-based speech enhancement has widely been studied and has largely surpassed the traditional noise estimation and Bayesian-filtering-based methods \cite{ephraim1984,cohen2001}, An overview of deep-learning-based speech enhancement methods can be found in \cite{wang2018}. These methods are often conducted in the time-frequency (TF) domain, and use a neural network to map noisy speech spectral feature onto clean speech targets. The input features, e.g. (logarithm) signal spectra, cepstral coefficients and linear prediction based features, generally represent the frame-wise full-band spectral structure of noisy speech. The output target consists of either the clean speech (logarithm) spectral vector or an TF mask vector to be applied to the corresponding noisy speech frame. Various training targets are summarized in \cite{wang2014}. Recurrent neural network (RNN), especially memory enhanced RNNs, such as long short-term memory (LSTM) \cite{hochreiter1997} and gated recurrent units (GRUs) \cite{cho2014}, are widely used for speech enhancement to model the temporal dynamics of the signal spectra.  \cite{weninger2014}, one of the first works using LSTM to perform speech enhancement, takes  the logarithmic Mel-scale spectrograms of noisy signal and clean speech as input and output, respectively. It was shown in \cite{chen2017} that LSTM outperforms the generic multilayer perceptron network. In \cite{sun2017}, multiple-target learning was proposed for LSTM based speech enhancement. A recent work \cite{xia2020} used GRU to improve the computational efficiency compared to LSTM. Instead of directly using signal features as input to RNNs, convolutional-recurrent neural networks \cite{tan2018convolutional,li2019convolutional} employ several convolutional layers prior to the RNN layer to extract more abstract spectral representations. All these RNN-based methods  process the frame-wise full-band spectra. When only forward RNN is used, these methods are naturally suitable for online processing.

In this work, we propose a speech enhancement method based on an output-delayed subband LSTM network. For each frequency, a sequence of STFT (short-time Fourier transform) magnitude of this frequency (together with some context frequencies) of a noisy signal input is fed into the LSTM network, which outputs the corresponding sequence of clean speech target at this frequency. A unique LSTM network is trained across all frequencies, and hence is shared by all frequencies during inference. With the aim of discriminating between speech and noise, this subband LSTM network is designed and expected to perform on the following grounds. First, it learns the frequency-wise signal stationarity to discriminate between speech and stationary noise. It is known that speech is non-stationary, while many types of noise are relatively stationary. The temporal evolution of frequency-wise STFT magnitude reflects the stationarity, which is the foundation for the conventional noise power estimators \cite{gerkmann2012,li2016icassp} and speech enhancement methods \cite{ephraim1984,cohen2001}. In our previous work \cite{li2019spl}, it was demonstrated that the subband LSTM network is able to be trained as a good noise power estimator for relatively stationary noise. In this work, we train the subband LSTM network to directly estimate clean speech. Second, it learns the local spectral pattern presented in the current and context frequencies to discriminate between speech and non-stationary/instantaneous noise. There was one attempt to use subband feature in \cite{wang2013} to perform TF-wise speech/noise classification, which testifies that local spectral pattern is informative for discriminating between speech and other signals. 

Exploiting a look-ahead is able to improve the performance. This causes a processing latency, but a small latency is normally tolerable for many applications. Bidirectional RNN is usually used for exploiting future information, which however is difficult to use for online processing. In this work, we adopt a simple yet effective network, i.e. output-delayed LSTM network \cite{turek2019}, to exploit the look-ahead with forward LSTM. When training the forward LSTM network, the output sequence is set to be delayed relative to the input sequence. To do this, the network learns to store the future information used for inferring one frame in the memory cell and hidden units of the forward LSTM. During inference, at one frame, the network receives the input vector of this frame, and predicts the speech target for one previous frame. Experiments show that this network achieves a speech enhancement performance close to the one of bidirectional LSTM.

%\subsection{Paper contributions}
Unlike the subband method of \cite{wang2013} and the aforementioned full-band methods, the proposed one shares the same network parameters across subbands, i.e. frequencies. This results in a drastic reduction of the number of network parameters as well as of the size of the training dataset. 
% Compared to the aforementioned full-band methods, by sharing the network across frequencies, the proposed network requires a much smaller training dataset, since which is already able to provide a large set of frequency-wise training samples. 
The proposed method focuses on the subband information and has a low-dimensional input vector, which makes the proposed network less subject to the \emph{curse of dimensionality} relative to the full-band networks. 
%As a result, the proposed network i) has a small size and with less trainable parameters; ii) is easy to train since the feature space can be more easily covered by the training data. 
The network training procedure converges normally in a few epochs, and overfitting rarely happens, which leads to a good generalization performance. In our previous work \cite{li2019waspaa}, we proposed to preform multichannel speech enhancement based on subband LSTM. In this work, we study the online single-channel (monaural) speech enhancement problem based on subband LSTM. 
We experimentally testify the effectiveness of subband LSTM for monaural speech enhancement. We exploit the subband spectral patterns to discriminate between speech and instantaneous noise, as \cite{li2019waspaa} only adopts the frequency-wise information. We develop the output-delayed LSTM network with look-ahead, for online speech enhancement.

\section{The Proposed Method}
\label{sec:method}

We consider single-channel signal in the STFT domain:
\begin{equation}
x(k,t)=s(k,t)+u(k,t), 
\end{equation}
where $k=0,\dots,K-1$ and $t=1,\dots,T$ denote the frequency and frame indices, respectively, $x(k,t)$, $s(k,t)$ and $u(k,t)$ are the (complex-valued) STFT coefficients of the microphone, speech and noise signals, respectively. The noise-free speech $s(k,t)$ represents the reverberant image signal received at the microphone.  This work focuses only on the denoising task, and the target is to suppress noise $u(k,t)$ and  recover the reverberant speech signal $s(k,t)$.

\subsection{Target and Input}
To recover the speech STFT coefficients $s(k,t)$, one popular way is to first predict a magnitude-based mask, such as ideal ratio mask or magnitude ratio mask \cite{wang2014}, and then reconstruct the complex-valued coefficients using the phase of noisy signal $x(k,t)$. In order to also estimate the phase of clean speech, \cite{williamson2016} proposed a complex ideal ratio mask (cIRM), defined as the ratio of the STFT coefficients between the speech and noisy signals. The cIRM was shown in \cite{williamson2016} in the full-band framework, and is verified by our preliminary experiments in our subband framework, to outperform the magnitude-based mask. In this work, we directly adopt cIRM as the training target, and compute cIRM exactly following the equations presented in \cite{williamson2016}. For one TF bin, we denote cIRM as $\mathbf{y}(k,t)\in \mathbb{R}^2$.  For each frequency bin, we aim to predict the cIRM sequence
\begin{equation}\label{eq:yseq}
\tilde{\mathbf{y}}(k) = \big(\mathbf{y}(k,1),\dots,\mathbf{y}(k,t),\dots,\mathbf{y}(k,T) \big).
\end{equation} 
Different from the full-band framework, e.g. \cite{wang2018}, (the network is trained to learn a regression from a full-band noisy spectrum to a full-band clean-speech target), in this work we devise a network to learn an speech-to-noise discrimination function based on signal stationarity and on the local spectral pattern, and then to predict the frequency-wise clean-speech target. For one TF bin, the STFT magnitude of the current frequency and the neighbor frequencies are concatenated to form the input, 
\begin{align}\label{eq:x}
\mathbf{x}(k,t) = [&|x(k,t)|,|x(k-1,t)|,\dots,|x(k-N,t)|, \nonumber \\ 
&|x(k+1,t)|,\dots,|x(k+N,t)|]^T \in \mathbb{R}^{2N+1},
\end{align}    
where $|\cdot|$ and $^T$ denote the absolute value and transpose operators, respectively, and $N$ is the number of neighbour frequencies considered on each side. For boundary frequencies, with $k-N<0$ or $k+N>K-1$, circular Fourier frequencies are used. For frequency $k$, the input sequence is then
\begin{align}\label{eq:xseq}
\tilde{\mathbf{x}}(k) = \big(\mathbf{x}(k,1),\dots,\mathbf{x}(k,t),\dots,\mathbf{x}(k,T) \big).
\end{align}  
In this sequence, the temporal evolution of $|x(k,t)|$ in \eqref{eq:x} reflects the signal stationarity, which is an efficient cue to discriminate between speech and relatively stationary noise. To some extend, the local spectral pattern encoded in $\mathbf{x}(k,t)$ and its temporal dynamics is able to discriminate between speech from other non-stationary or instantaneous noise signals.

As for training, input-target sequence pairs are generated with a constant-length sequence. To facilitate network training, the input sequence has to be normalized to equalize the input levels. We empirically normalize the input sequence with the mean STFT magnitude of the present frequency, i.e. $\frac{1}{T}\sum_{t=1}^{T}|x(k,t)|$.

\subsection{Output-Delayed LSTM Network}

RNN transmits the hidden units along the time steps. To avoid the problem of exponential weight decay (or explosion) along time, LSTM introduces an extra memory cell %which conveys the information along time step respectively to the hidden units. The memory cell 
allowing to learn long-term dependencies \cite{hochreiter1997}.

\begin{figure}[t!]
\centering

\pgfdeclarelayer{background}

\pgfsetlayers{background}

\usetikzlibrary{shapes.geometric,backgrounds,arrows,calc,fit}
\tikzstyle{io} = [rectangle, rounded corners, minimum width=4cm, minimum height=1.2cm, text centered, draw=black,align=center]
\tikzstyle{lstm} = [rectangle, minimum width=4cm, minimum height=1.2cm, text centered, draw=black,align=center]
\tikzstyle{blstm} = [rectangle,dashed, minimum width=4cm, minimum height=1.2cm, text centered, draw=black,align=center]
\tikzstyle{data1} = [rectangle, rounded corners, minimum width=2.4cm, minimum height=0.8cm, text centered, draw=black,align=center]
\tikzstyle{nobox} = [rectangle, rounded corners, draw=none,fill=white,minimum width=1cm, minimum height=0cm, text centered, align=center]
 \tikzstyle{constant} = [rectangle, minimum width=3.5cm, minimum height=1.1cm,text centered, draw=black,align=center] 
 
 \tikzstyle{arrow} = [->,>=latex]
 \tikzstyle{aarrow} = [->,>=latex,dashed]

\resizebox{0.99\columnwidth}{!}{
\begin{tikzpicture}

%% main flow
\large
\node (xt) [io] {{Input} $\mathbf{x}(t)$ \\ size: $2F+1$};

\node (t) [below of=xt,node distance=1.0cm] {\large $t$};
\node (d1) [right of=t,node distance=2.5cm] {$\cdots$};
\node (T) [right of=d1,node distance=1.5cm] {$T$};

\node (l1t) [lstm,above of =xt,node distance=1.5cm] {Forward LSTM \\ output size: $384$};
\node (l2t) [lstm,above of =l1t,node distance=1.5cm] {Forward LSTM \\ output size: $256$};
\node (denset) [lstm,above of =l2t,node distance=1.5cm] {Dense \\ output size: $2$};
\node (yt) [io,above of =denset,node distance=1.5cm] {{Output} $y(t-\tau)$ \\ size: $2$};

\node (xtm1) [io,left of =xt,node distance=5.0cm] {{Input} $\mathbf{x}(t-1)$ \\ size: $2F+1$};
\node (tm1) [below of=xtm1,node distance=1.0cm] {\large $t-1$};
\node (d2) [left of=tm1,node distance=2.5cm] {$\cdots$};
\node (t1) [left of=d2,node distance=1.5cm] {$1$};

\node (l1tm1) [lstm,above of =xtm1,node distance=1.5cm] {Forward LSTM \\ output size: $384$};
\node (l2tm1) [lstm,above of =l1tm1,node distance=1.5cm] {Forward LSTM \\ output size: $256$};
\node (densetm1) [lstm,above of =l2tm1,node distance=1.5cm] {Dense \\ output size: $2$};
\node (ytm1) [io,above of =densetm1,node distance=1.5cm] {{Output} $y(t-1-\tau)$ \\ size: $2$};

\draw [arrow]  (xt) --  (l1t);
\draw [arrow]  (l1t) --  (l2t);
\draw [arrow]  (l2t) --  (denset);
\draw [arrow] (denset) --  (yt);

\draw [arrow]  (xtm1) --  (l1tm1);
\draw [arrow]  (l1tm1) --  (l2tm1);
\draw [arrow]  (l2tm1) --  (densetm1);
\draw [arrow] (densetm1) --  (ytm1);

\draw [arrow] (l1tm1) --  (l1t);
\draw [arrow] (l2tm1) --  (l2t);

\node (l2tm2) [nobox, left of =l2tm1,node distance=3.7cm] {};
\draw [arrow]  (l2tm2) --  (l2tm1);
\node (l1tm2) [nobox, left of =l1tm1,node distance=3.7cm] {};
\draw [arrow]  (l1tm2) --  (l1tm1);
\node (l2tp1) [nobox, right of =l2t,node distance=3.7cm] {};
\draw [arrow]  (l2t) --  (l2tp1);
\node (l1tp1) [nobox, right of =l1t,node distance=3.7cm] {};
\draw [arrow]  (l1t) --  (l1tp1);

\end{tikzpicture}
}
\vspace{-0.0cm}
\caption{Diagram of the proposed output-delayed subband LSTM network.} 
\label{fig:lstm}
\vspace{-0.3cm}
\end{figure}
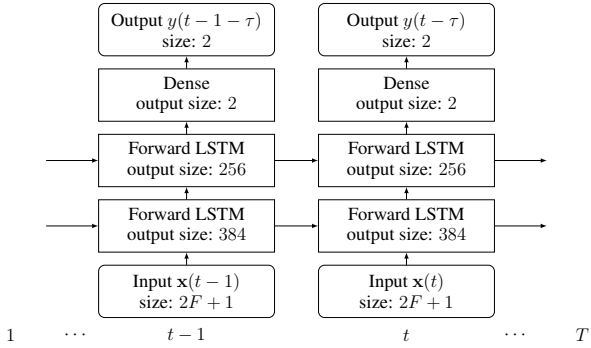

For online processing, the network receives and processes data one time step at a time. A look-ahead causes a processing latency. Many of the speech applications can tolerate tens to hundreds of milliseconds of latency.  In this work, we adopt a simple yet effective network, i.e. an output delayed LSTM network, to exploit the look-ahead advantages. Fig.~\ref{fig:lstm} shows the network diagram used in the present work. Two layers of unidirectional forward LSTM networks are stacked, followed by a dense layer to output the prediction. The network input is the noisy signal sequence defined in (\ref{eq:xseq}), while the network output is the cIRM sequence defined in (\ref{eq:yseq}), but with a delay $\tau$ relative to the input sequence. Note that the frequency index $k$ is omitted, since all the frequencies share the same network with the same parameters, and is equivalently taken as training/inference samples by the network. The LSTM layers are trained to learn a speech/noise discriminative function based on the signal stationarity and on the local spectral pattern, and then to predict the target. To infer $\mathbf{y}(t-\tau)$, the input of the future time steps, i.e. $\mathbf{x}(t-\tau+1),\dots,\mathbf{x}(t)$, are provided. To exploit the future time steps, instead of adopting extra backward LSTMs, this network stores future information, used for inferring $\mathbf{y}(t-\tau)$ in the memory cell and hidden units of the forward LSTM. At inference time, the network behaves the same as the standard LSTM, namely unrolling the forward LSTM once per time step, and automatically exploiting the future time steps $\tau$.   

The proposed output-delayed subband network is of relatively small in size, namely approximatively 1.3 millions learnable parameters, due to the fact that we focus on learning the subband information with a relatively low-dimensional input vector. The mean squared error (MSE), i.e. $|\mathbf{y}(k,t)-\hat{\mathbf{y}}(k,t)|^2$, is used as the training loss, where $\hat{\mathbf{y}}(k,t)$ represents the prediction.

%\end{document}%

\subsection{Online Inference}

During inference, the signal is received and processed frame by frame, and each frequency is processed independently and sequentially. First, the raw input signal is normalized using the mean of the STFT magnitude, which is recursively updated:
\begin{align}\label{eq:mu}
\mu(k,t) = \alpha\mu(k,t-1)+(1-\alpha)|x(k,t)|,
\end{align}
where $\alpha=\frac{L-1}{L+1}$ is a smoothing parameter, with $L$ being the desired number of frames to be smoothed. Then the network predicts the cIRM for bin $(k,t-\tau)$, using the normalized input vector together with the memory cell and hidden units from $t-1$, i.e. $c(k,t-1)$ and $h(k,t-1)$. Let $\text{LSTM}()$ denote one LSTM inference, formally we have:
\begin{align}\label{eq:online}
\hat{y}(k,t-\tau) = \text{LSTM}\Big(\frac{\mathbf{x}(k,t)}{\mu(k,t)},c(k,t-1),h(k,t-1)\Big).
\end{align}
Finally, the STFT coefficients of the enhanced speech are computed using the predicted cIRM following the equations presented in \cite{williamson2016}. 

\section{Experiments}
\label{sec:experiment}

\subsection{Experimental Setup}

We took part to the INTERSPEECH 2020 Deep Noise Suppression (DNS) challenge \cite{reddy2020} whose objective is real-time single-channel speech enhancement. The clean speech dataset for training is extracted from the audio book dataset Librivox by selecting good quality recordings, with more than 500 hours of speech from 2,150 speakers. The noise dataset consists of about 150 audio classes and 60,000 clips extracted from Audioset, and 10,000 noise clips from Freesound and DEMAND databases. 

\setlength{\tabcolsep}{5pt}
\begin{table}[t]
\centering
\caption{Speech enhancement results as a function of $N$ with $\tau=0$ (\emph{\textbf{up}}) and $\tau$ with $N=15$ (\emph{\textbf{bottom}}). The scores are averaged over the test sets \emph{no\_reverb} and \emph{with\_reverb}, with SNR of 0-20 dB. 'bi' denotes bidirectional LSTM.}
\label{tab:Ntau}
\begin{tabular}{c | c c c c c c }  
 $N$ & noisy & 7 & 9 &11 & 13 & 15 \\ \hline
 PESQ & 2.34  & 2.93  &  2.97  &  2.98  &  2.98  &  2.99    \\
 STOI (\%) & 89.1  & 91.3  &  91.7  &  91.7  &  91.7  &  91.7    \\
 SDR (dB) & 9.1   & 14.0  & 14.4   & 14.4   & 14.6   & 14.6    \\ 
\end{tabular}
\vspace{3mm}
\begin{tabular}{c | c c c c c c c }  
 $\tau$ & 0 & 2 & 4 & 6 & 8 & 10 & bi  \\ \hline
 PESQ & 2.99 &   3.10 &    3.12 &    3.13 &    3.16  &   3.13  &  3.20    \\
 STOI (\%) & 91.7  &  92.6 &   92.7  &  93.0    & 93.1   & 93.0 &    93.3    \\
 SDR (dB) & 14.6 &   15.4 &   15.4 &   15.5 &   15.6 &   15.7 &   16.5   \\
\end{tabular}
\vspace{-.3cm}
\end{table}

\begin{table}[t]
\centering
\caption{Brief comparison between NSNet and of the proposed network. }
\label{tab:network}
\begin{tabular}{c  | c | c  }  
  &NSNet  & prop. \\ \hline
Input dimension    & 771   &31 \\
Network  & 1 Dense (500) & 1 LSTM (384)  \\  
         & 3 GRUs (500) &  1 LSTM (256) \\
         & 1 Dense (514) &   1 Dense (2)\\
Output dimension  & 514   & 2 \\
\# Parameters  & 5.1 M & 1.3 M \\
Training data  & 500 hours & 20 hours \\
\# Training epochs  & $\approx$150 & $\approx$8 \\  
\end{tabular}
\vspace{-.3cm}
\end{table}

\begin{table*}[t]
\centering
\caption{Speech enhancement results obtained with NSNet and with the proposed method (with $\tau=2$). }
\vspace{-.2cm}
\label{tab:cmp}
\begin{tabular}{c  | c c c | ccc | ccc | ccc }  
 & \multicolumn{3}{c|}{no\_reverb, SNR 0-20 dB} & \multicolumn{3}{c|}{with\_reverb, SNR 0-20 dB} & \multicolumn{3}{c|}{no\_reverb, SNR 0 dB} & \multicolumn{3}{c}{with\_reverb, SNR 0 dB} \\
 & noisy & NSNet & prop. & noisy & NSNet & prop. & noisy & NSNet & prop. & noisy & NSNet & prop. \\ \hline
PESQ  &  2.16 & 2.67 & 2.84 & 2.52 & 2.94 & 3.26 & 1.56 & 1.93 & 2.19 & 1.75 & 2.10 & 2.40 \\
STOI (\%)  & 91.5 & 94.5 & 94.4 & 86.6 & 90.4 & 90.9 & 80.9 & 87.8 & 86.7 & 72.2 & 80.4 & 80.0\\
SDR (dB) & 9.1 & 15.9 & 15.8 & 9.2 & 15.3 & 15.1 & 0.0 & 10.6 & 9.2 & 0.2 & 9.8 & 8.7 \\
\end{tabular}
\vspace{-.4cm}
\end{table*}

\noindent
\textbf{Training data generation.}
To generate reverberant speech, two real-recorded multichannel room impulse response (RIR) datasets are used: (i)~the Multichannel Impulse Response Database \cite{hadad2014}, with three reverberation times (T60) 0.16~s, 0.36~s and 0.61~s, and (ii)~the Reverb challenge \cite{kinoshita2016} dataset with three reverberation times 0.3~s, 0.6~s and 0.7~s. The two datasets are recorded with two different eight-channel microphone arrays, with various microphone-to-speaker distances and speaker directions. Single-channel RIRs are randomly selected from these multichannel RIRs and convoluted with the clean speech clips to generate reverberant speech. Then the reverberant speech and randomly selected noise clips are added to generate noisy speech, with a signal-to-noise-ratio (SNR) randomly selected from the set of $\{-5,0,5,10,15\}$~dB. Using the properly modified DNS signal generation script, a total of 20 hours of noisy speech signals are generated for training, from which 5 hours of data are reverberation-free speech. The signals are transformed to the STFT domain using a 512-sample (32~ms) Hanning window with a frame step of 256 samples. The sequence length for training is set to $T=192$ frames (about 3~s). The training sequences are picked out from the clip-level signals with 50\% overlap for two adjacent sequences. In total, about 11.1 million training sequences are generated. It is obvious that the amount of training data, i.e. 20 hours, is way less than the amount provided in the DNS dataset, which means that only a small part of the DNS data are used. Such a small amount of training data is already able to provide a rich set of subband spectral patterns. 

\noindent
\textbf{Test data.}
The DNS challenge provides a publicly available test set including two categories of synthetic clips, i.e. without and with reverberations. Each category has 150 noisy clips with SNR levels distributed between 0~dB to 20~dB. In addition to these data, for each category, we generate one new group of noisier speech by remixing the test clean speech and noise, all with SNR of 0 dB. 

\noindent
\textbf{Implementation details.}
We use Keras \cite{chollet2015keras} to implement the proposed method. The Adam optimizer \cite{kingma2014adam} is used with a learning rate of 0.001. The batch size is set to 512. The training sequences were shuffled. The number of train epochs is empirically set to 8. For speech enhancement, one clip of test data is processed frame by frame. The online inference is implemented using the stateful LSTM to transmit the memory cell and hidden units. The function  'predict\_on\_batch' is used to process all frequencies. The smoothing factor $\alpha$ is set with $L=192$.

\subsection{Experimental Results}

Three performance metrics are used, i) perceptual evaluation of speech quality (PESQ) \cite{rix2001};  ii) short-time objective intelligibility (STOI) \cite{taal2011}; and iii) signal-to-distortion ratio (SDR) \cite{vincent2006} in dB. For all metrics, the larger the better. \footnote{Please visit \url{https://team.inria.fr/perception/research/onlinese-lstm/} for subjective test on audio clips.}

\noindent
\textbf{Setting the number of context frequencies.} To predict the mask for one single frequency, $N$ neighbour frequencies for each side are adopted as the network input to exploit the subband spectral pattern that is able to discriminate between speech and noise. Table \ref{tab:Ntau} (\textbf{up}) shows the performance as a function of $N$. It is seen that, with the increasing of $N$, the performance measures increase and converge until $N=15$. This means, in the present framework, the network needs $2N+1=31$ frequencies (about 969 Hz) to fully exploit the subband spectral pattern. $N$ is set to 15 for the following experiments. 

\noindent
\textbf{Experiments with output delay.} The network output is set as a delayed sequence relative to the input sequence. This allows the unidirectional forward LSTM to exploit future information, and to facilitate the online processing with a look-ahead. Table \ref{tab:Ntau} (\textbf{bottom}) shows the performance as a function of $\tau$. It is seen that the performance measures noticeably increase with increasing $\tau$, until $\tau=8$. When a larger $\tau$ is set, e.g. $\tau=10$, the performance measures decrease, which is possibly due to the fact that more look-ahead frames do not provide more useful information, but increase the training difficulty. The last column in this table presents the results of bidirectional LSTM which is only suitable for offline applications. It is unsurprising that the performance measures of the output delayed LSTM are lower the ones of bidirectional LSTM, but the performance gap is small, especially for PESQ and STOI. This indicates that even if the output-delayed LSTM can not fully exploit the future information, it is still very efficient. 

\noindent
\textbf{Comparison with the DNS baseline method.}
The DNS challenge baseline method, i.e. NSNet \cite{xia2020}, is similar with the proposed method in the sense that both methods use RNN to perform the so-called single-frame-in and single-frame-out online speech enhancement. The main difference is that NSNet takes as input the full-band noisy spectra and outputs the full-band speech mask, while the proposed network takes as input the subband noisy spectra and outputs the frequency-wise speech mask. Thence NSNet provides a perfect comparison method that can demonstrate the difference between the full-band and subband frameworks.  We use the \emph{dns\_challenge} recipe provided in the asteroid toolkit \footnote{\url{https://github.com/mpariente/asteroid/tree/master/egs/dns_challenge}} to implement NSNet. The configurations provided in the recipe are used, except that the training data are generated with reverberations and SNRs following the principles used for the proposed method. A complex network is trained, which takes as input the complex and magnitude spectra and output the complex mask. Table \ref{tab:network} briefly summarizes NSNet and the proposed network. Since NSNet learns the full-band spectra, it requires a larger network and more training resources than the proposed subband network. 

According to the rule of the DNS challenge that a maximum of 40 ms look-ahead can be used, we submitted enhanced signals yielded by the proposed method with $\tau=2$, which exploits two future frames to enhance the current frame, and uses a $16\times 2=32$~ms look-ahead. Table \ref{tab:cmp} shows the performance of NSNet and of our submissions. The comparison between NSNet and the proposed method is quite consistent across the four different reverberant and noisy conditions. NSNet slightly outperforms the proposed method in terms of SDR and STOI, and the superiority is more noticeable for the low SNR cases (with SNR=0 dB). This indicates that the full-band spectra is indeed more discriminative than the subband spectra in the aspect of suppressing the noise level. The proposed method yields higher PESQ scores than NSNet by a large margin. The reasons for this could be that (i)~the proposed method discriminates between speech and noise highly relying on the signal stationarity, and thus is especially superior to suppress relatively stationary/continuous noise, and that (ii)~NSNet predicts the speech mask of all frequencies together, for which the prediction errors are correlated and structured along frequencies, which leads to noticeable speech or noise distortion. By contrast, the proposed method processes frequencies independently, hence the prediction errors are likely to be independent as well. By some informal listening test, the enhanced signals of the proposed method sound smoother and more natural than NSNet. Moreover, the suppressed noise is rarely distorted by the proposed method. By  the listening test, the proposed method suppresses the noise level and does not introduce other unpleasant sound effects. 

%\noindent
%\textbf{Computation time.} 
Inference was run on an Intel i5-1035G1 quad core CPU with a base frequency of 1.0~GHz. The computation of per STFT frame takes 7.0~ms, which is in real-time.
The DNS Challenge has announced the P.808 subjective evaluation results on a blind noisy test set (consists of both synthetic and real data). The Mean Opinion Scores (MOS) of the noisy set is 3.01, and the one of the proposed method is 3.32, which ranks the 4th place out of the 16 Real-Time track submissions. 
%Audio examples are available in our website. \footnote{https://team.inria.fr/perception/research/mse-lstm/}

\section{Conclusion}
\label{sec4}

We proposed an online monaural speech enhancement method based on a delayed subband LSTM network. Focusing on one frequency at a time, the proposed network requires a small amount of training data and of resources. Promising speech denoising results were achieved, which testifies that subband information, i.e. signal stationarity and local spectral patterns, is indeed discriminative for the task at hand. The output-delayed scheme provides a computationally efficient yet performant way to exploit the look-ahead potential.

% -------------------------------------------------------------------------
% Either list references using the bibliography style file IEEEtran.bst
\bibliographystyle{IEEEtran}
%\bibliography{lixf_bib}

\begin{thebibliography}{10}
\providecommand{\url}[1]{#1}
\csname url@samestyle\endcsname
\providecommand{\newblock}{\relax}
\providecommand{\bibinfo}[2]{#2}
\providecommand{\BIBentrySTDinterwordspacing}{\spaceskip=0pt\relax}
\providecommand{\BIBentryALTinterwordstretchfactor}{4}
\providecommand{\BIBentryALTinterwordspacing}{\spaceskip=\fontdimen2\font plus
\BIBentryALTinterwordstretchfactor\fontdimen3\font minus
  \fontdimen4\font\relax}
\providecommand{\BIBforeignlanguage}[2]{{%
\expandafter\ifx\csname l@#1\endcsname\relax
\typeout{** WARNING: IEEEtran.bst: No hyphenation pattern has been}%
\typeout{** loaded for the language `#1'. Using the pattern for}%
\typeout{** the default language instead.}%
\else
\language=\csname l@#1\endcsname
\fi
#2}}
\providecommand{\BIBdecl}{\relax}
\BIBdecl

\bibitem{ephraim1984}
Y.~Ephraim and D.~Malah, ``Speech enhancement using a minimum-mean square error
  short-time spectral amplitude estimator,'' \emph{IEEE Transactions on
  Acoustics, Speech and Signal Processing}, vol.~32, no.~6, pp. 1109--1121,
  1984.

\bibitem{cohen2001}
I.~Cohen and B.~Berdugo, ``Speech enhancement for non-stationary noise
  environments,'' \emph{Signal processing}, vol.~81, no.~11, pp. 2403--2418,
  2001.

\bibitem{wang2018}
D.~Wang and J.~Chen, ``Supervised speech separation based on deep learning: An
  overview,'' \emph{IEEE/ACM Transactions on Audio, Speech, and Language
  Processing}, vol.~26, no.~10, pp. 1702--1726, 2018.

\bibitem{wang2014}
Y.~Wang, A.~Narayanan, and D.~Wang, ``On training targets for supervised speech
  separation,'' \emph{IEEE/ACM Transactions on Audio, Speech, and Language
  Processing}, vol.~22, no.~12, pp. 1849--1858, 2014.

\bibitem{hochreiter1997}
S.~Hochreiter and J.~Schmidhuber, ``Long short-term memory,'' \emph{Neural
  computation}, vol.~9, no.~8, pp. 1735--1780, 1997.

\bibitem{cho2014}
K.~Cho, B.~Van~Merri{\"e}nboer, C.~Gulcehre, D.~Bahdanau, F.~Bougares,
  H.~Schwenk, and Y.~Bengio, ``Learning phrase representations using {RNN}
  encoder-decoder for statistical machine translation,''
  \emph{arXiv:1406.1078}, 2014.

\bibitem{weninger2014}
F.~Weninger, F.~Eyben, and B.~Schuller, ``Single-channel speech separation with
  memory-enhanced recurrent neural networks,'' in \emph{IEEE International
  Conference on Acoustics, Speech and Signal Processing (ICASSP)}, 2014, pp.
  3709--3713.

\bibitem{chen2017}
J.~Chen and D.~Wang, ``Long short-term memory for speaker generalization in
  supervised speech separation,'' \emph{The Journal of the Acoustical Society
  of America}, vol. 141, no.~6, pp. 4705--4714, 2017.

\bibitem{sun2017}
L.~Sun, J.~Du, L.-R. Dai, and C.-H. Lee, ``Multiple-target deep learning for
  lstm-rnn based speech enhancement,'' in \emph{2017 Hands-free Speech
  Communications and Microphone Arrays}.\hskip 1em plus 0.5em minus 0.4em\relax
  IEEE, 2017, pp. 136--140.

\bibitem{xia2020}
Y.~{Xia}, S.~{Braun}, C.~K.~A. {Reddy}, H.~{Dubey}, R.~{Cutler}, and
  I.~{Tashev}, ``Weighted speech distortion losses for neural-network-based
  real-time speech enhancement,'' in \emph{IEEE International Conference on
  Acoustics, Speech and Signal Processing}, 2020, pp. 871--875.

\bibitem{tan2018convolutional}
K.~Tan and D.~Wang, ``A convolutional recurrent neural network for real-time
  speech enhancement.'' in \emph{Interspeech}, vol. 2018, 2018, pp. 3229--3233.

\bibitem{li2019convolutional}
A.~Li, C.~Zheng, and X.~Li, ``Convolutional recurrent neural network based
  progressive learning for monaural speech enhancement,'' \emph{arXiv preprint
  arXiv:1908.10768}, 2019.

\bibitem{gerkmann2012}
T.~Gerkmann and R.~C. Hendriks, ``Unbiased {MMSE}-based noise power estimation
  with low complexity and low tracking delay,'' \emph{IEEE Transactions on
  Audio, Speech, and Language Processing}, vol.~20, no.~4, pp. 1383--1393,
  2012.

\bibitem{li2016icassp}
X.~Li, L.~Girin, S.~Gannot, and R.~Horaud, ``Non-stationary noise power
  spectral density estimation based on regional statistics,'' in \emph{IEEE
  International Conference on Acoustics, Speech and Signal Processing}, 2016,
  pp. 181--185.

\bibitem{li2019spl}
X.~Li, S.~Leglaive, L.~Girin, and R.~Horaud, ``Audio-noise power spectral
  density estimation using long short-term memory,'' \emph{IEEE Signal
  Processing Letters}, 2019.

\bibitem{wang2013}
Y.~Wang, K.~Han, and D.~Wang, ``Exploring monaural features for
  classification-based speech segregation,'' \emph{IEEE Transactions on Audio,
  Speech, and Language Processing}, vol.~21, no.~2, pp. 270--279, 2013.

\bibitem{turek2019}
J.~S. Turek, S.~Jain, M.~Capota, A.~G. Huth, and T.~L. Willke, ``{A
  single-layer RNN can approximate stacked and bidirectional RNNs, and
  topologies in between},'' \emph{arXiv:1909.00021}, 2019.

\bibitem{li2019waspaa}
X.~Li and R.~Horaud, ``Multichannel speech enhancement based on time-frequency
  masking using subband long short-term memory,'' in \emph{IEEE Workshop on
  Applications of Signal Processing to Audio and Acoustics}, 2019, pp.
  298--302.

\bibitem{williamson2016}
D.~S. Williamson, Y.~Wang, and D.~Wang, ``Complex ratio masking for monaural
  speech separation,'' \emph{IEEE/ACM Transactions on Audio, Speech, and
  Language Processing}, vol.~24, no.~3, pp. 483--492, 2016.

\bibitem{reddy2020}
C.~K.~A. Reddy, E.~Beyrami, H.~Dubey, V.~Gopal, R.~Cheng, R.~Cutler,
  S.~Matusevych, R.~Aichner, A.~Aazami, S.~Braun, P.~Rana, S.~Srinivasan, and
  J.~Gehrke, ``The {Interspeech} 2020 deep noise suppression challenge:
  Datasets, subjective speech quality and testing framework,'' 2020.

\bibitem{hadad2014}
E.~Hadad, F.~Heese, P.~Vary, and S.~Gannot, ``Multichannel audio database in
  various acoustic environments,'' in \emph{IEEE International Workshop on
  Acoustic Signal Enhancement}, 2014, pp. 313--317.

\bibitem{kinoshita2016}
K.~Kinoshita, M.~Delcroix, S.~Gannot, E.~A. Habets, R.~Haeb-Umbach,
  W.~Kellermann, V.~Leutnant, R.~Maas, T.~Nakatani, B.~Raj \emph{et~al.}, ``A
  summary of the reverb challenge: state-of-the-art and remaining challenges in
  reverberant speech processing research,'' \emph{EURASIP Journal on Advances
  in Signal Processing}, vol. 2016, no.~1, p.~7, 2016.

\bibitem{chollet2015keras}
F.~Chollet \emph{et~al.}, ``Keras,'' \url{https://github.com/fchollet/keras},
  2015.

\bibitem{kingma2014adam}
D.~P. Kingma and J.~Ba, ``Adam: A method for stochastic optimization,'' in
  \emph{International Conference on Learning Representations}, 2015.

\bibitem{rix2001}
A.~W. Rix, J.~G. Beerends, M.~P. Hollier, and A.~P. Hekstra, ``Perceptual
  evaluation of speech quality ({PESQ})-a new method for speech quality
  assessment of telephone networks and codecs,'' in \emph{IEEE International
  Conference on Acoustics, Speech, and Signal Processing}, vol.~2, 2001, pp.
  749--752.

\bibitem{taal2011}
C.~H. Taal, R.~C. Hendriks, R.~Heusdens, and J.~Jensen, ``An algorithm for
  intelligibility prediction of time--frequency weighted noisy speech,''
  \emph{IEEE Transactions on Audio, Speech, and Language Processing}, vol.~19,
  no.~7, pp. 2125--2136, 2011.

\bibitem{vincent2006}
E.~Vincent, R.~Gribonval, and C.~F{\'e}votte, ``Performance measurement in
  blind audio source separation,'' \emph{IEEE Transactions on Audio, Speech,
  and Language Processing}, vol.~14, no.~4, pp. 1462--1469, 2006.

\end{thebibliography}


\end{document}